\documentclass[aps, prd, amsmath, amssymb, floats, floatfix, twocolumn, nofootinbib, superscriptaddress]{revtex4-1}
\usepackage{graphicx}
\usepackage{color}
\usepackage{latexsym}
\usepackage{amsfonts}
\usepackage{dcolumn}
\usepackage{bm}

\newcommand{\be}{\begin{eqnarray}}
\newcommand{\ee}{\end{eqnarray}}

\let\sb=_ \catcode`\_=\active \def_#1{\ensuremath \sb{\rm#1}}

\begin{document}


\title{A general study of regular and singular black hole solutions\\in Einstein's conformal gravity}

\author{Qiqi~Zhang}
\affiliation{Department of Physics, Southern University of Science and Technology (SUSTech), Shenzhen 518055, China}

\author{Leonardo~Modesto}
\email[Corresponding author: ]{lmodesto@sustc.edu.cn}
\affiliation{Department of Physics, Southern University of Science and Technology (SUSTech), Shenzhen 518055, China}

\author{Cosimo~Bambi}
\affiliation{Center for Field Theory and Particle Physics and Department of Physics, Fudan University, 200438 Shanghai, China}
\affiliation{Theoretical Astrophysics, Eberhard-Karls Universit\"at T\"ubingen, 72076 T\"ubingen, Germany}

\date{\today}

\begin{abstract}
We study some general properties of two black hole solutions in Einstein's conformal gravity. Both solutions can be obtained from the Kerr metric with a suitable conformal rescaling, which leads, respectively, to a regular and a singular spacetime. In addition to the mass $M$ and the spin angular momentum $J$ of the black hole, these solutions are characterized by a new parameter, $L$, which may be expected to be of the order of the black hole mass. We study the geodesic motion and we calculate the radiative efficiency of a putative accretion disk around these black holes, which allows us to get an estimate of an upper bound on the value of $L$. Lastly, we explore the possibility of ``destroying'' the event horizon of these black holes. Within our approach, we are not able to destroy the event horizon of the regular black hole solution, while we can in the case of the singular one.
\end{abstract}

\maketitle

\section{\label{sec:level1}Introduction}

Conformal invariance is an appealing proposal to solve spacetime singularities in Einstein's gravity~\cite{englert1976conformal,naelikar1977space,t2012quantum,mannheim2012making}. In conformal gravity, the theory is invariant under a conformal transformation of the metric tensor $g_{\mu\nu}$
\begin{equation}\label{eq-ct}
g_{\mu\nu} \to \hat{g}_{\mu\nu} = S \, g_{\mu\nu} \, ,
\end{equation}
where $S = S (x)$ is a function of the spacetime point. Conformal gravity can solve the problem of spacetime singularities (before and after the symmetry breaking) by finding a suitable conformal transformation $S$  that removes the singularity and by interpreting the metric $\hat{g}_{\mu\nu}$ as the physical metric of the spacetime.

Einstein's gravity is not conformally invariant, but it can be made conformally invariant, for instance by introducing an auxiliary field~\cite{englert1976conformal,naelikar1977space,t2012quantum,mannheim2012making}. An examples of conformal theory of gravity in four dimensions is~\cite{dirac}
\begin{equation}
\begin{aligned}
\mathcal{L}=\phi^{2}R+6g^{\mu\nu}(\partial_{\mu}\phi)(\partial_{\nu}\phi),
\end{aligned}
\end{equation}
where $\phi$ is an auxiliary scalar field (dilaton). In our case, we are not interested in a particular model, but we just require the theory to be invariant under the transformation~(\ref{eq-ct})\footnote{We note that there are many different conformal gravity models in the literature. The first conformal extension of general relativity was introduced by Weyl in 1918~\cite{weyl1918gravitation}. More recently, Weyl's conformal gravity was revised by Mannheim and collaborators~\cite{mannheim2006alternatives}. Mannheim and Kazanas also found the fourth-order conformal Weyl gravity solution to the Kerr problem in ~\cite{mannheim1991solutions}, and these solutions were reconsidered in recent years by others in~\cite{varieschi2014kerr,mureika2017black}.}. While the theory is invariant under conformal transformations, a Higgs-like mechanism may choose one of the metric as the ``physical" solution to describe the spacetime~\cite{bambi2017spacetime}. The world around us is not conformally invariant and therefore, if conformal invariance is a fundamental symmetry of the spacetime, it should be broken, and one of the possibilities is that it is spontaneously broken.

In Ref.~\cite{bambi2017spacetime}, two of us found a singularity-free rotating black hole solution conformally equivalent to the Kerr metric. In Boyer-Lindquist coordinates, the line element reads
\begin{equation}\label{eq-reg}
d\hat{s}^{2}= \left( 1+\frac{L^{2}}{\Sigma} \right)^{4} ds^{2}_{\rm Kerr}
\end{equation}
where $\Sigma = r^{2}+a^{2} \cos^{2}\theta$ and $L>0$ is a new parameter with dimensions of a length. It is natural to expect that $L$ is either of the order of the Planck length, $L \sim L_{\rm Pl}$, or of the order of the black hole mass, $L \sim M$, because these are the only two scales already present in the model. In this paper, we consider the second scenario with $L$ of the order of $M$, because it is the only one with observational implications for astrophysical black holes. $ds^{2}_{\rm Kerr}$ is the line element of the Kerr metric
\be
ds^{2}_{\rm Kerr} &=&
-\left(1-\frac{2Mr}{\Sigma} \right) dt^{2}
-\frac{4Mar\sin^{2}\theta}{\Sigma} \, dt \, d\phi \nonumber\\
&& + \left(r^{2}+a^{2}+\frac{2Ma^{2}r\mathrm{sin}^{2}\theta}{\Sigma} \right)
\mathrm{sin}^{2}\theta \, d\phi^{2} \nonumber\\
&& +\frac{\Sigma}{\Delta} \, dr^{2}+\Sigma \, d\theta^{2} \, ,
\ee
$a=J/M$ is the  rotational parameter (the dimensionless spin parameter is $a_{*} = a/M$) and $\Delta = r^{2}-2Mr+a^{2}$.

In the present paper, we are not interested in the singularity problem, and we want to study the properties of rescaled metrics with different conformal factors. So we also include another conformally rescaled black hole solution which is still singular after the transformation and whose line element reads\footnote{In general, the exponent $n$ in $\left( 1 - L^2/\Sigma\right)^n$ may be any integer number. As discussed in Section~\ref{s-5}, within our study we find that we can destroy the black hole for $n$ odd and we cannot for $n$ even. To simplify the calculations, we chose 3, but the same qualitative results should hold for other odd numbers.}
\begin{equation}\label{eq-sing}
d\hat{s}^{2}= \left(1-\frac{L^{2}}{\Sigma} \right)^{3}ds^{2}_{Kerr} \, .
\end{equation}

Both the metrics in Eqs.~(\ref{eq-reg}) and (\ref{eq-sing}) are black hole solutions in the large family of Einstein's conformal theories of gravity. In the symmetric phase, these two solutions are physically equivalent, because the theory is invariant under conformal transformations and the two metrics in Eqs.~(\ref{eq-reg}) and (\ref{eq-sing}) only differ by the conformal factor. In the broken phase, different conformal factors produce different observable effects. Here we want to study the metrics in Eqs.~(\ref{eq-reg}) and (\ref{eq-sing}) as the two prototypes of a regular and a singular black hole solution, respectively, but there is not specific reason to choose them and not others.

The content of the paper is as follows. In Section~\ref{s-2}, we prove the regularity or singularity of the rescaled spacetimes. In Section~\ref{s-3}, we study the geodesic motion in these spacetimes, focusing our attention to equatorial circular orbits. In Section~\ref{s-c}, we check the possible existence of a Carter-like constant. In Section~\ref{s-4}, we calculate the radiative efficiency of a putative thin accretion disk around these black holes as a function of $L$, obtaining a constraint on the value of $L$. Finally, in Section~\ref{s-5}, we study the possibility of ``destroying'' the event horizon of these black holes by overspinning the object. With our set-up, we are not able to destroy the event horizon of the regular black holes, while we can in the case of the singular black holes. Our conclusions are reported in Section~\ref{s-6}. Throughout the paper, we employ geometrized units in which $G_{N} = c = 1$ and adopt a metric with signature $(- + + +)$.

\section{\label{s-2} geodesic completion (incompletion)}

Generally speaking, spacetime singularities are regions of a spacetime with some pathological properties. In this section, we check the geodesic completion/incompletion of photon orbits in the spacetimes with the line elements in Eqs.~(\ref{eq-reg}) and (\ref{eq-sing}). The geodesic completion for time-like and null geodesics for the metric in Eq.~(\ref{eq-reg}) was already proven in Ref.~\cite{bambi2017spacetime}, but it is convenient to summarize here to see the difference with the solution in Eq.~(\ref{eq-sing}). We confirm that the solution in Eq.~(\ref{eq-reg}) is regular, because photon can never reach the black hole center with a finite value of the affine parameter, while we find that the solution in Eq.~(\ref{eq-sing}) is singular.

\subsection{\label{sec:citeref}Regular black hole spacetime}

Let us first study the solution in Eq.~(\ref{eq-reg}), which can be derived from the Kerr metric after imposing the following rescaling factor
\begin{equation} \label{eq-scaling}
S = \left(1+\frac{L^{2}}{\Sigma} \right)^4 \, .
\end{equation}
Note that the Kretschmann scalar of the solution in Eq.~(\ref{eq-reg}) has the form
\begin{equation}
\hat{\mathcal K} = 
\frac{1}{\left( \Sigma + L^2 \right)^n} \times 
\left(\text{polynomial in }r,x,M,a,L\right) \, ,
\end{equation}
where $x=\cos\theta$ and $n$ is an integer number. This expression is everywhere regular for $L \neq 0$; that is, $\hat{\mathcal K}$ never diverges. For $L=0$, we recover the well-known result of the Kerr metric in which the Kretschmann scalar diverges when $r \rightarrow 0$ for $\theta = \pi/2$.

In order to study the geodesic completion of the spacetime, we proceed as follows. The spacetime is stationary and axisymmetric, which leads to have the following two Killing vectors
\begin{equation}
\xi^{\alpha} = (1,0,0,0) \, , \quad
\eta^{\alpha} = (0,0,0,1) \, ,
\end{equation}
and the following conserved quantities
\be\label{eq-e}
e &=& -\xi^{\alpha}u^{\beta}\hat{g}_{\alpha\beta} = - (\hat{g}_{tt}u^{t}+\hat{g}_{t\phi}u^{\phi}) \, , \\
\label{eq-l}
\ell &=& \eta^{\alpha}u^{\beta}\hat{g}_{\alpha\beta} = \hat{g}_{\phi t}u^{t}+\hat{g}_{\phi \phi}u^{\phi} \, .
\ee
For photons, the conservation of mass reads
\be\label{eq-mass}
&& \hat{g}_{\alpha\beta}u^{\alpha}u^{\beta} = 0 \nonumber\\
&& \Rightarrow  \hat{g}_{tt}\dot{t}^{2}+2\hat{g}_{t\phi}\dot{t}\dot{\phi}
+\hat{g}_{rr}\dot{r}^{2}+\hat{g}_{\theta\theta}\dot{\theta}^{2}+\hat{g}_{\phi \phi}\dot{\phi}^{2} = 0 \, ,
\ee
where we used the notation $\dot{x}^\mu = u^\mu$. From Eqs.~(\ref{eq-e}) and (\ref{eq-l}), we can write $\dot{t}$ and $\dot{\phi}$ in terms of $e$, $\ell$, and the metric coefficients
\begin{equation}
\dot{t} = \frac{e\hat{g}_{\phi \phi}+\hat{g}_{\phi t}\ell}{\hat{g}_{\phi t}^{2}-\hat{g}_{tt}\hat{g}_{\phi \phi}} \, , \quad
\dot{\phi} = \frac{e\hat{g}_{\phi t}+\hat{g}_{tt}\ell}{\hat{g}_{\phi t}^{2}-\hat{g}_{tt}\hat{g}_{\phi \phi}} \, .
\end{equation}
If we plug these expressions of $\dot{t}$ and $\dot{\phi}$ into Eq.~(\ref{eq-mass}), we obtain
\begin{equation}\label{eq-veff}
\hat{g}_{rr}\dot{r}^{2}
+ \hat{g}_{\theta\theta}\dot{\theta}^{2}
+\frac{e^{2}\hat{g}_{\phi \phi}+2e\hat{g}_{\phi t}\ell+\hat{g}_{tt}\ell^{2}}{\hat{g}_{tt}\hat{g}_{\phi \phi}-\hat{g}_{\phi t}^{2}} = 0 \, .
\end{equation}

\begin{figure*}
\begin{minipage}{0.48\linewidth}
\centerline{\includegraphics[width=8.5cm]{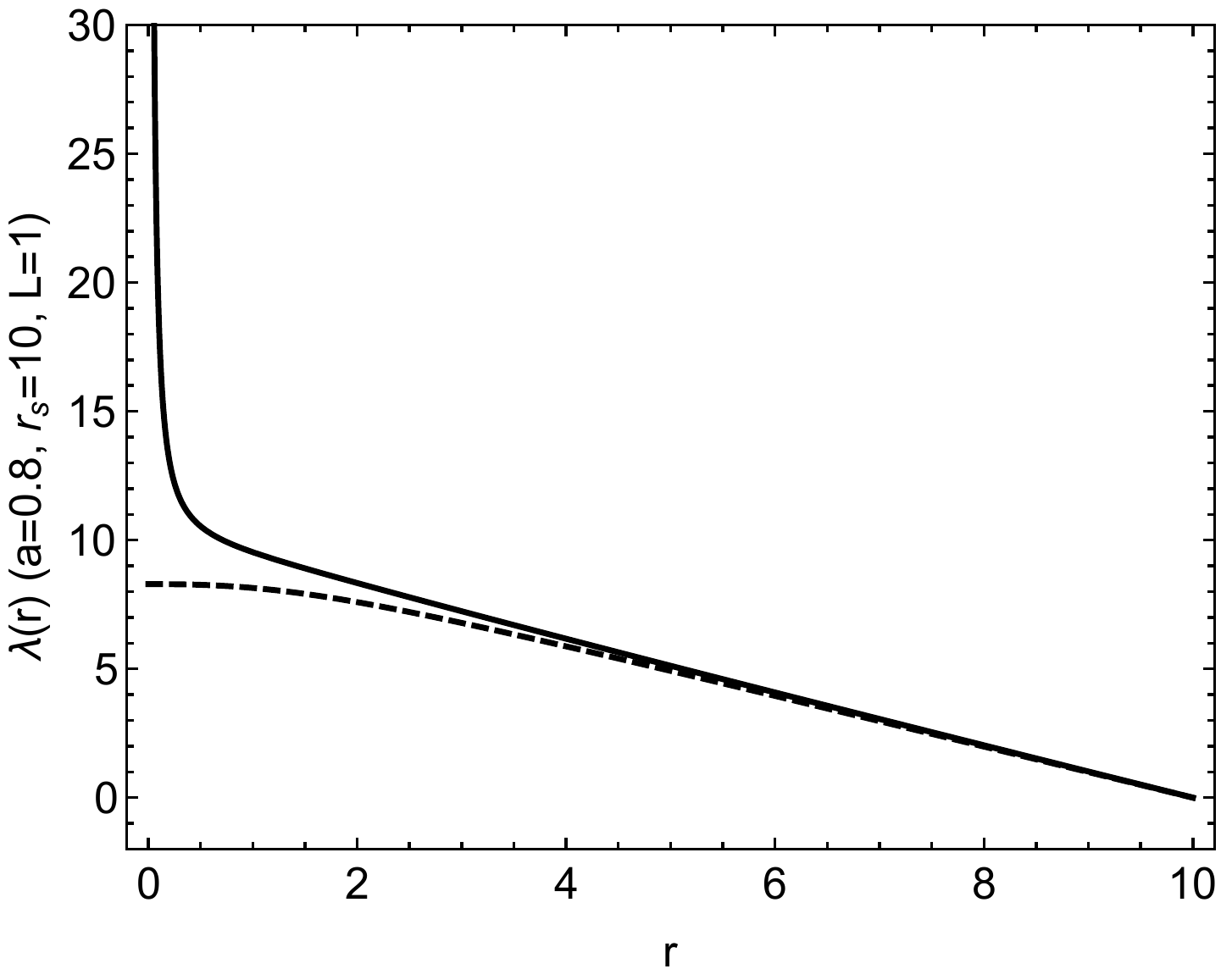}}
\centerline{(a)}
\end{minipage}
\hfill
\begin{minipage}{0.48\linewidth}
\centerline{\includegraphics[width=8.5cm]{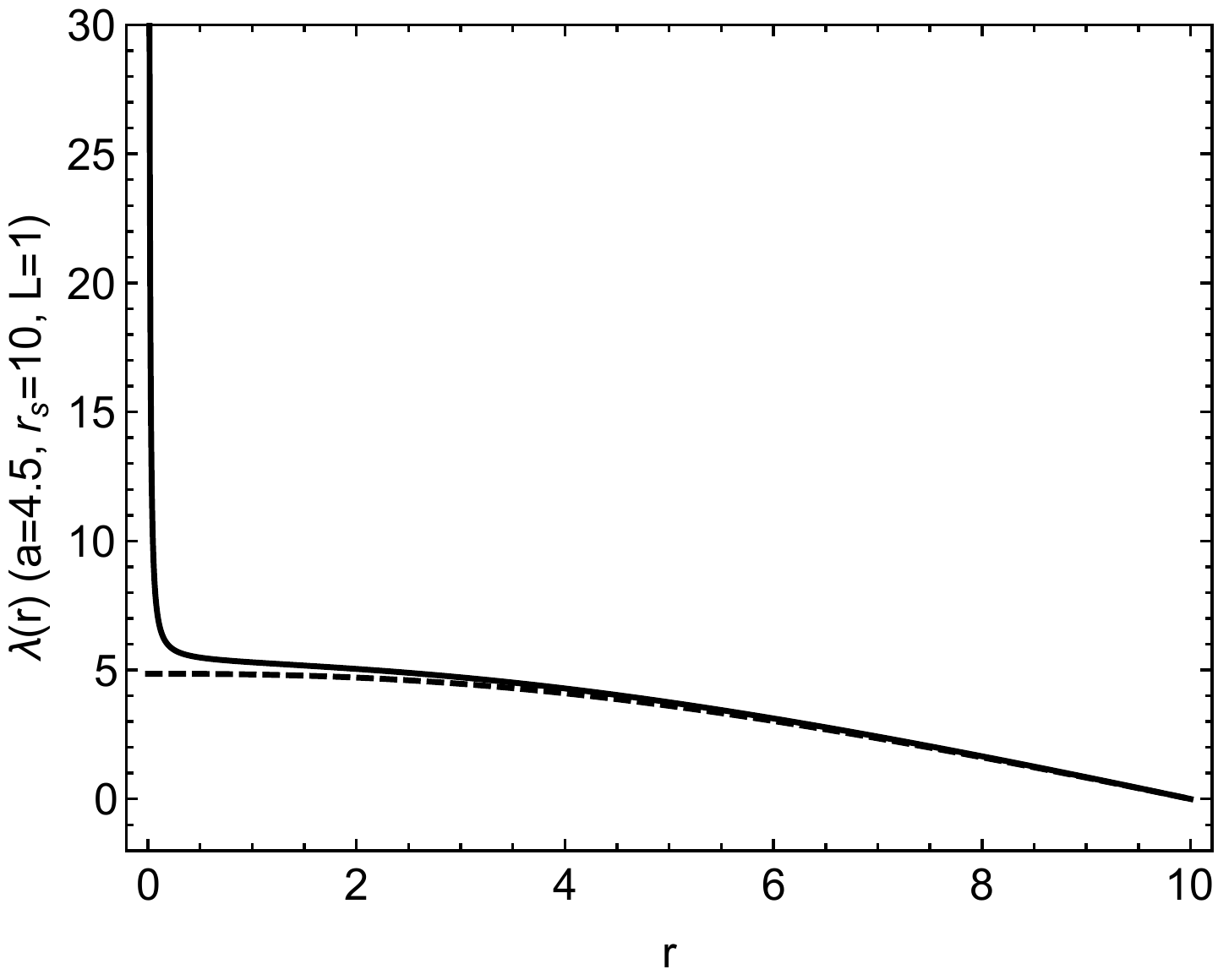}}
\centerline{(b)}
\end{minipage}
\vfill
\vspace{0.8cm}
\begin{minipage}{0.48\linewidth}
\centerline{\includegraphics[width=8.5cm]{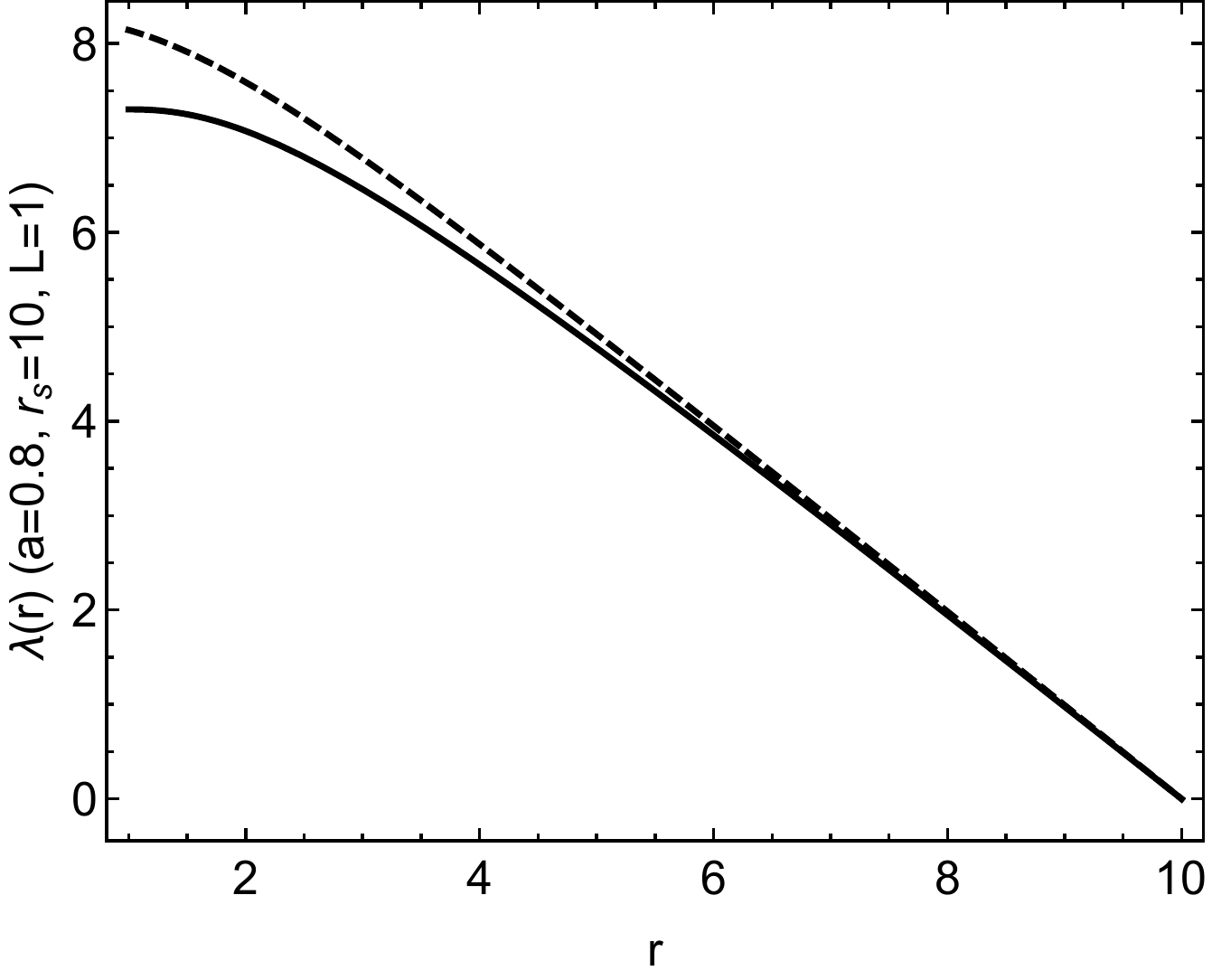}}
\centerline{(c)}
\end{minipage}
\hfill
\begin{minipage}{0.48\linewidth}
\centerline{\includegraphics[width=8.5cm]{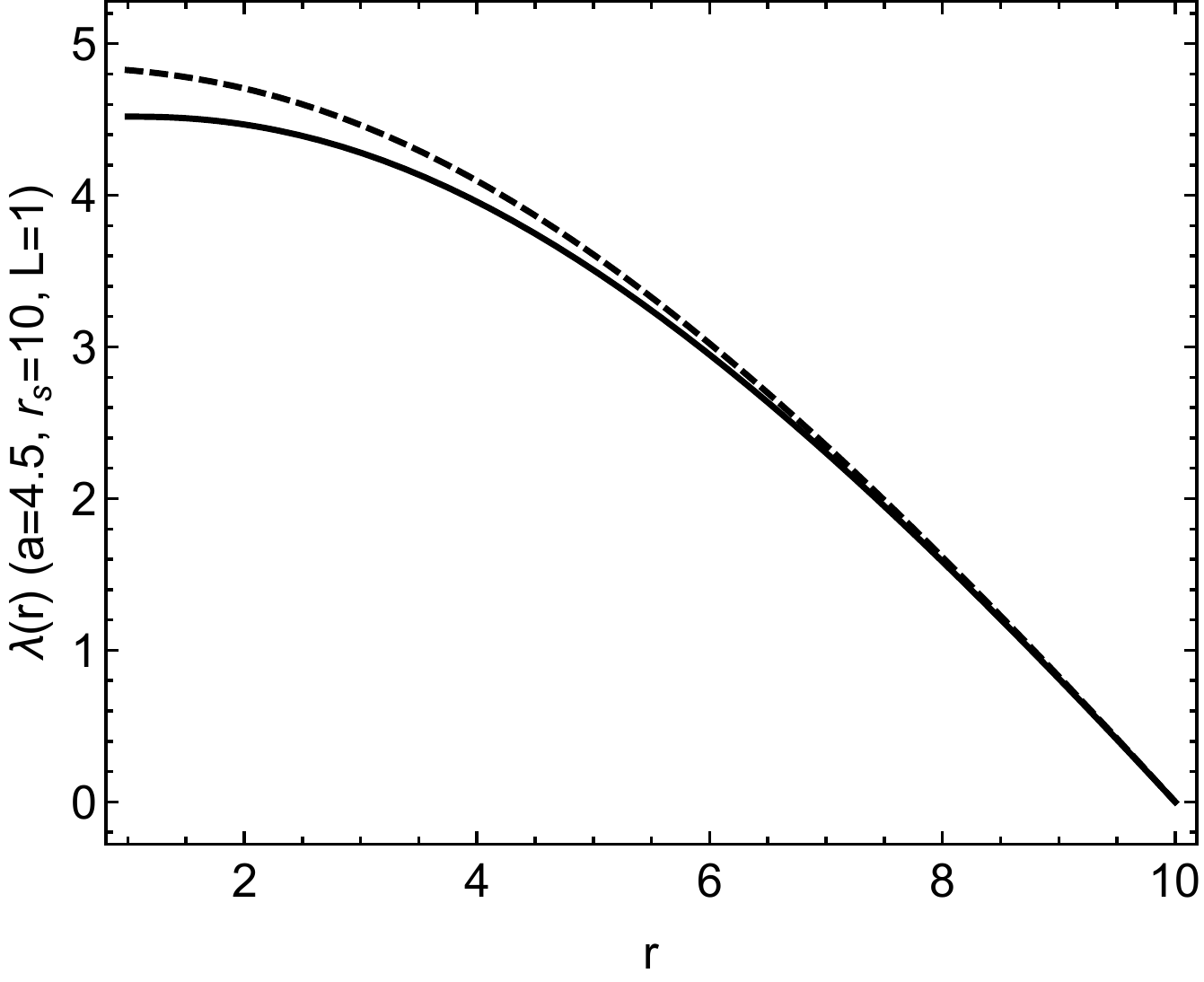}}
\centerline{(d)}
\end{minipage}
\vspace{0.2cm}
\caption{Here we employ the following values for the parameters and the conserved quantities: $L = 1$, $r_{in} = 10$, $a = 0.8$, $\theta = \pi/2$, and $r_{s} = 10$ for panels (a) and (c), $L = 1$, $r_{in} = 10$, $a = 4.5$, $\theta = \pi/2$, and $r_{s} = 10$ for panels (b) and (d). The dashed line shows the situation for the standard Kerr metric. (a) Plot of the affine parameter $\lambda (r)$ for an infalling massless particle in the regular black hole spacetime. The affine parameter $\lambda\to+\infty$ for $r\to+0$, and therefore the spacetime is geodesically complete.  (b) As in panel~(a) but for $a = 4.5$. (c) Plot of the affine parameter $\lambda(r)$ for an infalling massless particle in the singular black hole spacetime. The affine parameter is finite when $r=1$. (d) As in panel~(c) but for $a = 4.5$. \label{fig-geo} }
\end{figure*}

For simplicity, now we restrict the attention to the motion in the equatorial plane, $\theta=\pi/2$ and $\dot{\theta}=0$, with vanishing angular momentum, $\ell = 0$ (which does not mean vanishing angular frequency for $\hat{g}_{\phi \rm t} \neq 0$). Eq.~(\ref{eq-veff}) becomes
\begin{equation}\label{eq-veff2}
\hat{g}_{rr}\dot{r}^{2}+\frac{e^{2}\hat{g}_{\phi \phi}}{\hat{g}_{tt}\hat{g}_{\phi \phi}-\hat{g}_{\phi t}^{2}} = 0 \, .
\end{equation}
For a rescaled Kerr metric in Boyer-Lindquist coordinates, namely $\hat{g}_{\mu\nu} = S \,  g_{\mu\nu}^{\rm Kerr}$ where $g_{\mu\nu}^{\rm Kerr}$ is the Kerr metric in Boyer-Lindquist coordinates, Eq.~(\ref{eq-veff2}) becomes
\begin{equation}
\frac{r^{3} S^2}{r^{3}+ra^{2}+r_{s}a^{2}} \left(\frac{dr}{d\lambda}\right)^{2} = e^{2} \, ,
\end{equation}
where $r_{\rm s} = 2M$.
This expression can be integrated by parts to obtain the affine parameter $\lambda (r)$
\begin{equation}
\lambda (r) = -e^{2} \int_{r_{in}}^{r} \sqrt{\frac{r^{3} S^2}{r^{3}+ra^{2}+r_{s}a^{2}}} \, dr \, , 
\end{equation}
where $r_{\rm in}$ is the initial value of the radial coordinate and the minus sign in front of the integral is because we are considering a photons moving from larger to smaller radii. Panels (a) and (b) in Fig.~\ref{fig-geo} show, for two particular cases for $S$ in Eq.~(\ref{eq-scaling}), that $\lambda\to+\infty$ as $r\to 0$; that is, the spacetime is geodesically complete because null geodesics never reach the center at $r=0$.

\subsection{\label{sec:citeref}Singular black hole spacetime}

Let us now move to the spacetime described by the line element in Eq.~(\ref{eq-sing}). The metric can be derived from the Kerr solution with the following rescaling factor
\begin{equation} 
S = \left(1-\frac{L^{2}}{\Sigma} \right)^3 \, .
\end{equation}
The Kretschmann scalar of this spacetime has the form
\begin{equation}
\hat{\mathcal K} = 
\frac{1}{\left( \Sigma - L^2 \right)^m} \times 
\left(\text{polynomial in }r,x,M,a,L\right) \, ,
\end{equation}
where $m$ is an integer number, and is singular when $r = \sqrt{L^2 - a^2 \cos^2\theta}$; that is, there is a singular surface with a finite value of the radial coordinate.

We can proceed as in the previous case and study the null geodesics in the equatorial plane with vanishing angular momentum. Panels (c) and (d) in Fig.~\ref{fig-geo} show, for two particular cases, that $\lambda$ remains finite for $r \to L$. The spacetime is geodesically incomplete.

\section{\label{s-3} Equatorial circular orbits}

In this section, we study the geodesics in the equatorial plane for the two black hole solutions and for different values of $L$. Equatorial circular orbits are of particular interest because they are the orbits of the particles of the gas in thin accretion disks around a black hole~\cite{bambi2017black,review}.


The geodesic motion of a (massive) test-particle in a spacetime with metric $g_{\mu\nu}$ is governed by the Lagrangian
\begin{equation}
\mathcal{L}=\frac{1}{2}g_{\mu\nu}\dot{x}^{\mu}\dot{x}^{\nu},
\end{equation}
where $\dot{} =d/d\tau$ and $\tau$ is the particle proper time.

Let us consider a stationary and axisymmetric spacetime whose line element can be written in the following form
\begin{equation}
ds^{2}=g_{tt}dt^{2}+2g_{t\phi}dtd\phi+g_{rr}dr^{2}+g_{\theta\theta}d\theta^{2}+g_{\phi\phi}d\phi^{2} \, ,
\end{equation}
with the metric coefficients independent of the coordinates $t$ and $\phi$. Since the metric is independent of the coordinates $t$ and $\phi$, we have two constants of motion, namely the specific energy at infinity $E$ and the axial component of the specific angular momentum at infinity $L_{z}$:
\begin{equation}
p_{t}=g_{tt}\dot t + g_{t\phi}\dot \phi= -E
\end{equation}
\begin{equation}
p_{\phi}=g_{t\phi}\dot t + g_{\phi\phi}\dot \phi= L_{z}
\end{equation}
With the conservation of the rest-mass, $g_{\mu\nu}\dot x^{\mu}\dot x^{\nu}=-1$, we can write the equation
\begin{equation}\label{eq--effe-pot-0}
g_{rr}\dot r^{2} + g_{\theta\theta}\dot \theta^{2}=V_{eff}
\end{equation}
where the effective potential $V_{eff}$ is given by
\begin{equation}
V_{eff}=\frac{E^{2}g_{\phi\phi}+2EL_{z}g_{t\phi}+L_{z}^{2}g_{tt}}{g_{t\phi}^{2}-g_{tt}g_{\phi\phi}}-1 \, .
\end{equation}

Circular orbits in the equatorial plane are located at the zeros and the turning points of the effective  potential\footnote{This is equivalent to say that circular orbits in the equatorial plane are at a minimum of the effective potential and note that the minimum is zero, as follows from the fact the left hand side in Eq.~(\ref{eq--effe-pot-0}) is non-negative.}: $\dot r=\dot \theta=0$, which implies $V_{eff}=0$, and $\ddot r=\ddot \theta=0$, requiring, respectively, $\partial_{r}V_{eff}=0$ and $\partial_{\theta}V_{eff}=0$. From these conditions, one can obtain the angular velocity $\Omega$, $E$, and $L_{z}$ of the test-particle
\begin{equation}
\Omega_{\pm}=\frac{d\phi}{dt}=\frac{-\partial_{r}g_{t\phi}\pm\sqrt{(\partial_{r}g_{t\phi})^{2}-(\partial_{r}g_{tt})(\partial_{r}g_{\phi\phi})}}{\partial_{r}g_{\phi\phi}}
\end{equation}
\begin{equation}
E=-\frac{g_{tt}+g_{t\phi}\Omega}{\sqrt{-g_{tt}-2g_{t\phi}\Omega-g_{\phi\phi}\Omega^{2}}}
\end{equation}
\begin{equation}
L_{z}=\frac{g_{t\phi}+g_{\phi\phi}\Omega}{\sqrt{-g_{tt}-2g_{t\phi}\Omega-g_{\phi\phi}\Omega^{2}}}
\end{equation}
where the $``+$" sign is for corotating orbits and the $``-"$ sign for counterrotating ones.

$E$ and $L_{z}$ diverge when their denominator vanishes. This happens at the radius of the so-called photon orbit $r_{\gamma}$
\begin{equation}
g_{tt}+2g_{t\phi}\Omega+g_{\phi\phi}\Omega^{2}=0 \Rightarrow r=r_{\gamma}
\end{equation}
The radius of the marginally bound orbit $r_{mb}$ is defined by
\begin{equation}
E=-\frac{g_{tt}+g_{t\phi}\Omega}{\sqrt{-g_{tt}-2g_{t\phi}\Omega-g_{\phi\phi}\Omega^{2}}}=1 \Rightarrow r=r_{mb}
\end{equation}
The orbit is marginally bound, which means that the test-particle has the sufficient energy to escape to infinity.
The radius of the marginally stable orbit $r_{ms}$, more often called the ISCO radius $r_{ISCO}$, is defined by
\begin{equation}
\partial_{r}^{2}V_{eff}=0\  \mathrm{or}\  \partial_{\theta}^{2}V_{eff}=0 \Rightarrow r=r_{ISCO}
\end{equation}

\begin{figure*}
\begin{minipage}{0.48\linewidth}
\centerline{\includegraphics[width=8.5cm]{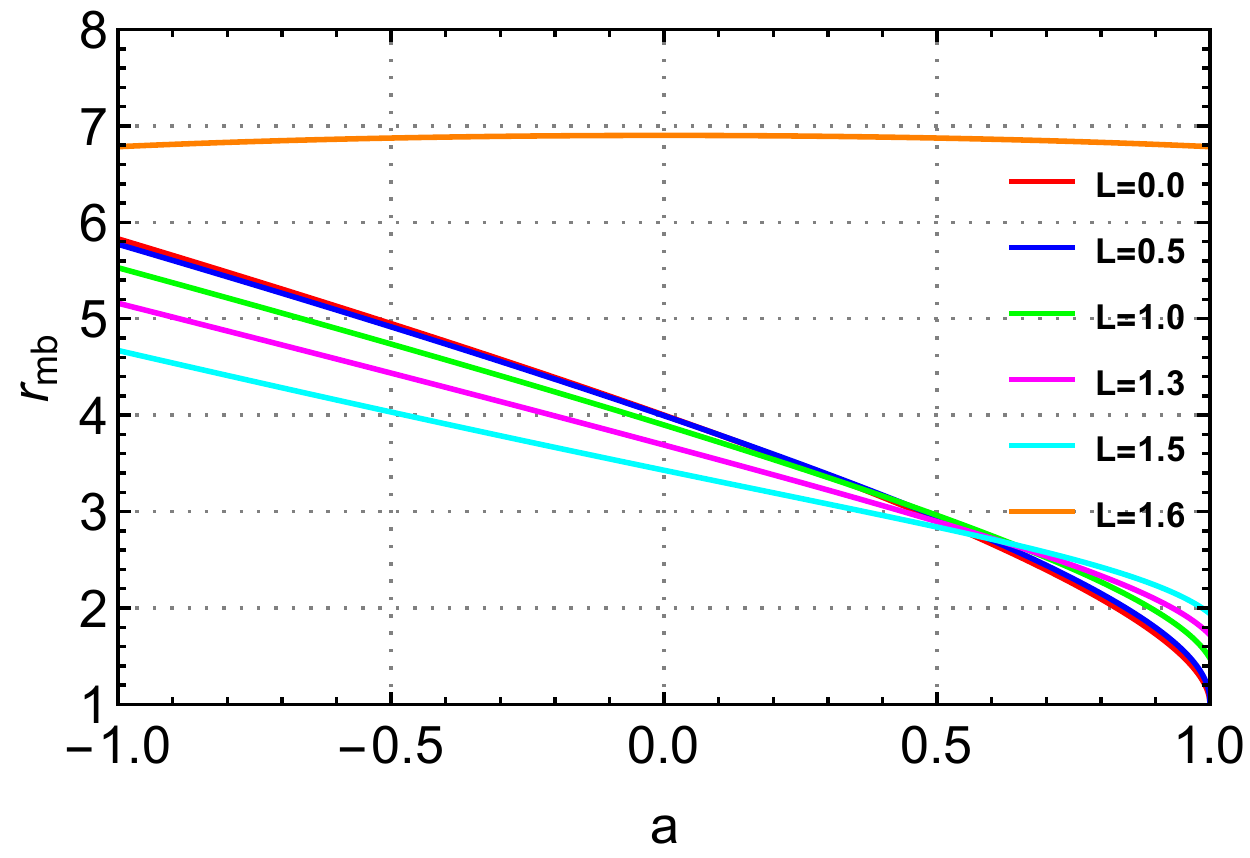}}
\centerline{(a)}
\end{minipage}
\hfill
\begin{minipage}{.48\linewidth}
\centerline{\includegraphics[width=8.5cm]{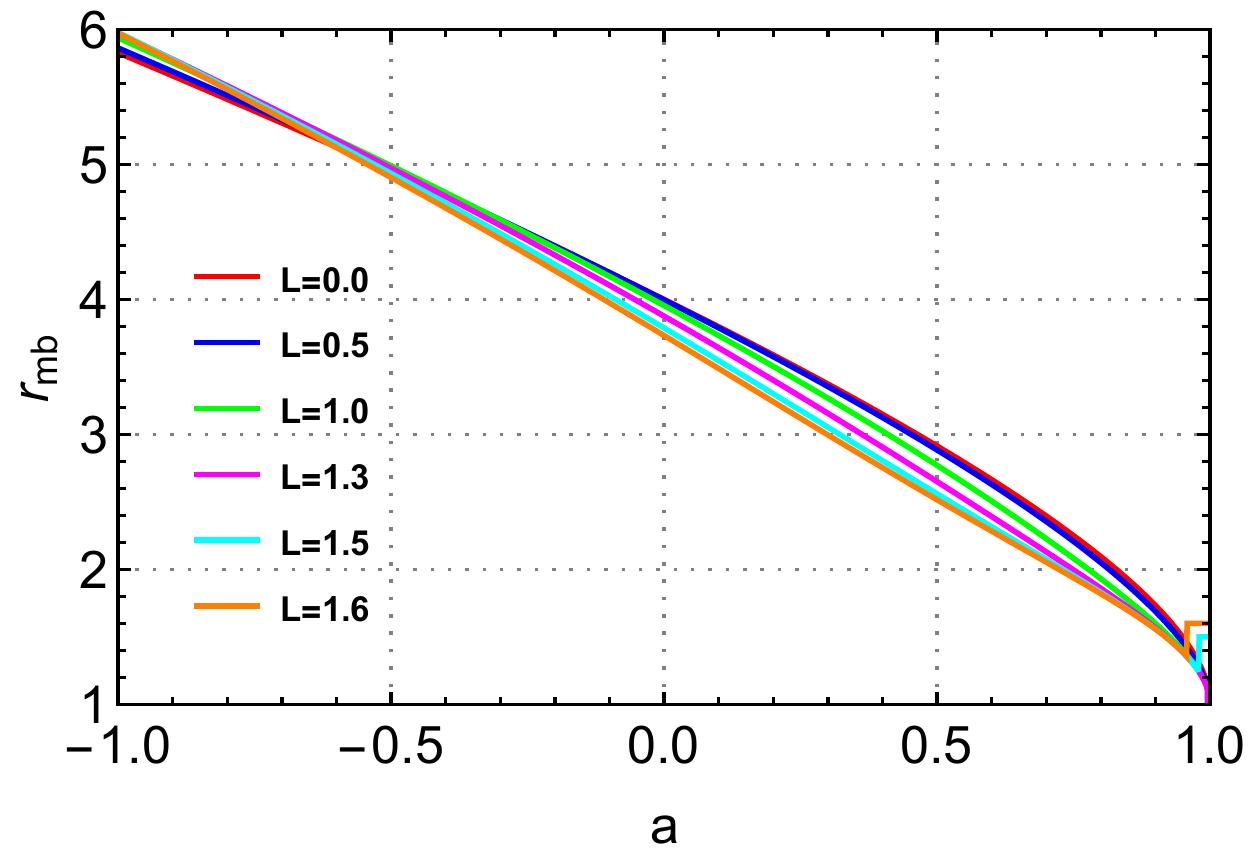}}
\centerline{(b)}
\end{minipage}
\vfill
\vspace{0.8cm}
\begin{minipage}{0.48\linewidth}
\centerline{\includegraphics[width=8.5cm]{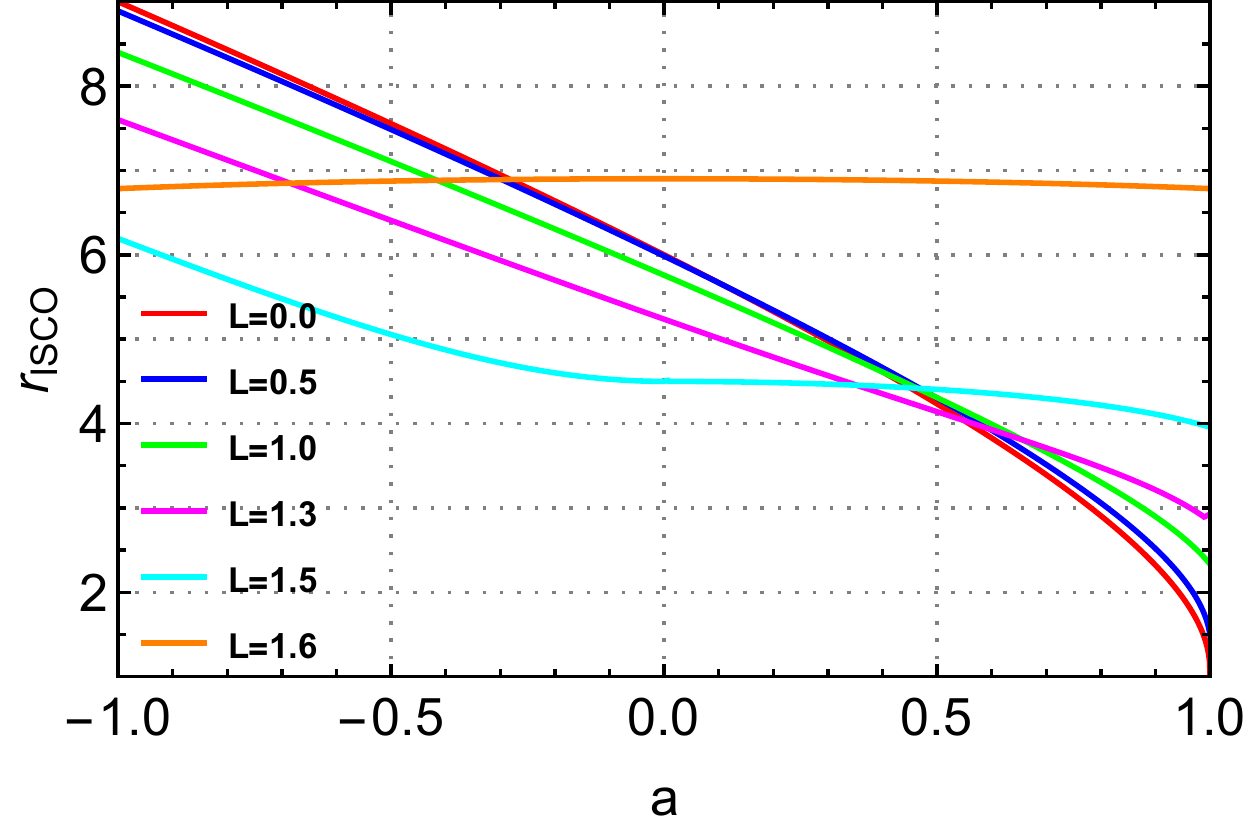}}
\centerline{(c)}
\end{minipage}
\hfill
\begin{minipage}{0.48\linewidth}
\centerline{\includegraphics[width=8.5cm]{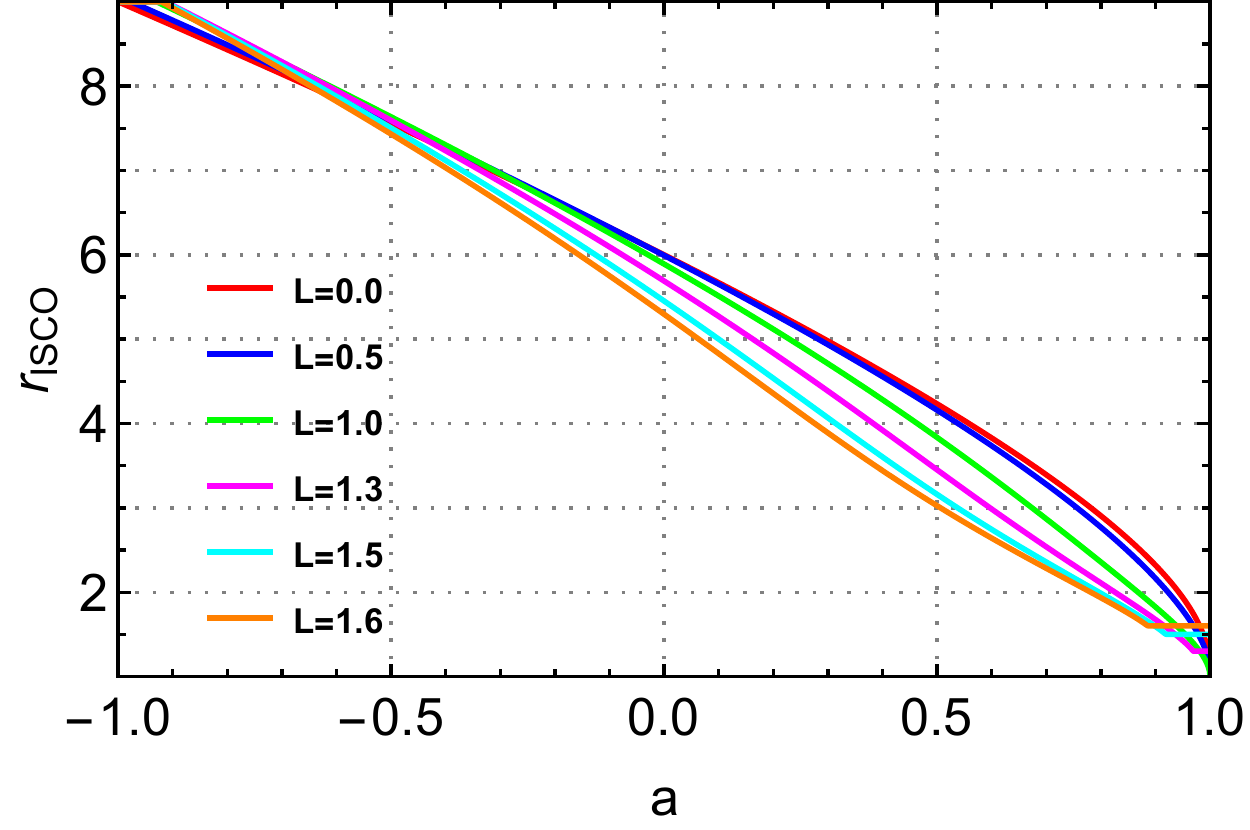}}
\centerline{(d)}
\end{minipage}
\vspace{0.2cm}
\caption{(a) Radius of marginally bound orbits as a function of the black hole spin and for $L$ varying from 0 to 1.6 in the regular black hole spacetimes. (b) As in panel (a) for the singular black hole solutions. (c) ISCO radius as a function of the black hole spin and for $L$ varying from 0 to 1.6 in the regular black hole spacetimes. (d) As in panel (c) for the singular black hole solutions. In these plots we assume units in which $M = 1$. \label{f-oo}}
\end{figure*}

Note that the photon orbit is independent of the rescaling factor $S$, and therefore our black hole solutions have the same photon orbits as in the Kerr metric for the same $M$ and $a$~\cite{bpt}
\begin{equation}
r_{\gamma}=2M \left\{1+\mathrm{cos} \left[\frac{2}{3}\mathrm{arccos}
\left(\mp\frac{a}{M}\right)\right] \right\} \, .
\end{equation}
On the contrary, the radius of the marginally bound orbit $r_{mb}$ and the ISCO radius $r_{ISCO}$ do depend on the scaling factor and $L$. Fig.~\ref{f-oo} show $r_{mb}$ and $r_{ISCO}$ as a function of the black hole spin for different values of $L$.

In the Kerr metric in Boyer-Lindquist coordinates, the radius of the event horizon is defined by the larger root of $g^{rr}=0$, which is equivalent to $\Delta=0$, and is
\begin{equation}\label{eq-r+}
r_{H}=M + \sqrt{M^{2}-a^{2}} \, .
\end{equation}
Since null trajectories are independent of the scaling factor, the event horizon in our regular and singular black hole spacetimes is still given by Eq.~(\ref{eq-r+}).

It is well-known that in the Kerr metric in Boyer-Lindquist coordinates 
\begin{equation}
r_H , \, r_\gamma , \, r_{mb} , \, r_{ISCO} \to M 
\end{equation}
for $a \to M$. However, this is an artifact of the Boyer-Lindquist coordinates, which are ill-defined at the event horizon, and these radii do not coincide~\cite{bpt}. In our regular and singular black hole spacetimes, for $a \to M$ we have still that $r_H , \, r_\gamma \to M$, because $r_H$ and $r_\gamma$ are the same as in the Kerr metric. However, the proper distance between $r_H$ and $r_\gamma$ depends on $L$ and it is thus different from the result in the Kerr solution. For the regular solution, for $a \to M$ the proper distance between $r_H$ and $r_\gamma$ becomes 
\begin{equation}
\int_{r_{H}}^{r_{\gamma}} \left(1+\frac{L^{2}}{r'^{2}}\right)^{2}\frac{r' dr'}{\sqrt{\Delta}} \rightarrow \left(L^{2}+M^{2} \right)^{2}\frac{\ln3}{2M^{3}} \, .
\end{equation}
For the singular solution, we find
\begin{equation}
\int_{r_{H}}^{r_{\gamma}} \left(1-\frac{L^{2}}{r'^{2}}\right)^{3/2}\frac{r' dr'}{\sqrt{\Delta}} \rightarrow -\left(L^{2}-M^{2}\right)^{3/2}\frac{\ln3}{2M^{2}} \, .
\end{equation}

\section{\label{s-c} Carter-like constant}

In general, the motion of a test-particle in a stationary and axisymmetric spacetime is characterized by three constants of motion: the particle mass, the energy, and the axial component of the angular momentum. The Kerr metric is a Petrov type~D spacetime, and therefore we can find a forth constant of motion, the so-called Carter constant~\cite{carter}. In certain coordinate systems, the presence of the Carter constant can simplify the calculations by separating the equations of motion~\cite{chandra}.

The Hamilton-Jacobi equation for the geodesic motion of a test-particle is
\begin{equation}\label{eq-ham}
2\frac{\partial {\mathcal S}}{\partial \tau}=g^{\mu\nu}\frac{\partial {\mathcal S}}{\partial x^{\mu}}\frac{\partial {\mathcal S}}{\partial x^{\nu}}
\end{equation}
where ${\mathcal S}$ is the Hamilton's principle function. The Kerr metric in Boyer-Lindquist coordinates is given by
\be
\left(\frac{\partial}{\partial s}\right)^{2}&=&-\frac{A}{\Sigma\Delta}\left(\frac{\partial}{\partial t}\right)^{2}-\frac{4aMr}{\Sigma\Delta}\left(\frac{\partial}{\partial t}\right)\left(\frac{\partial}{\partial \phi}\right)
\nonumber\\
&&+\frac{\Delta}{\Sigma}\left(\frac{\partial}{\partial r}\right)^{2}+\frac{1}{\Sigma}\left(\frac{\partial}{\partial \theta}\right)^{2}
\ee
where $A=(r^{2}+a^{2})^{2}-a^{2}\Delta \mathrm{sin}^{2}\theta$. Eq.~(\ref{eq-ham}) becomes
\begin{equation}\label{eq-ss}
\begin{aligned}
2\frac{\partial {\mathcal S}}{\partial \tau}=&-\frac{A}{\Sigma\Delta}\left(\frac{\partial {\mathcal S}}{\partial t}\right)^{2}-\frac{4aMr}{\Sigma\Delta}\left(\frac{\partial {\mathcal {\mathcal S}}}{\partial t}\right)\left(\frac{\partial S}{\partial \phi}\right)\\&+\frac{\Delta}{\Sigma}\left(\frac{\partial {\mathcal S}}{\partial r}\right)^{2}+\frac{1}{\Sigma}\left(\frac{\partial {\mathcal S}}{\partial \theta}\right)^{2}\\=&-\frac{1}{\Sigma\Delta}\left[\left(r^{2}+a^{2}\right)\frac{\partial {\mathcal S}}{\partial t}+a\frac{\partial {\mathcal S}}{\partial \phi}\right]^{2}\\&+\frac{1}{\Sigma \mathrm{sin}^{2}\theta}\left[a\mathrm{sin}^{2}\theta\frac{\partial {\mathcal S}}{\partial t}+\frac{\partial {\mathcal S}}{\partial \phi}\right]^{2}\\&+\frac{\Delta}{\Sigma}\left(\frac{\partial {\mathcal S}}{\partial r}\right)^{2}+\frac{1}{\Sigma}\left(\frac{\partial {\mathcal S}}{\partial \theta}\right)^{2} \, .
\end{aligned}
\end{equation}
We look for a solution of the Hamilton-Jacobi equation of the following form
\begin{equation}
{\mathcal S}=\frac{1}{2}\delta\tau-Et+L_{z}\phi+{\mathcal S}_{r}(r)+{\mathcal S}_{\theta}(\theta)
\end{equation}
where $\delta = -1$ for massive particles and $\delta = 0$ for photons. ${\mathcal S}_{r}$ and ${\mathcal S}_{\theta}$ are, respectively, functions of $r$ and $\theta$ only. Eq.~(\ref{eq-ss}) becomes
\begin{equation}
\begin{aligned}
\delta\Sigma&=\frac{1}{\Delta}\left[\left(r^{2}+a^{2}\right)E-aL_{z}\right]^{2}-\frac{1}{\mathrm{sin}^{2}{\theta}}\left(aE\mathrm{sin}^{2}{\theta}-L_{z}\right)^{2}\\&-\Delta\left(\frac{\partial {\mathcal S}}{\partial r}\right)^{2}-\left(\frac{\partial {\mathcal S}}{\partial \theta}\right)^{2}\\&=\frac{1}{\Delta}\left[\left(r^{2}+a^{2}\right)E-aL_{z}\right]^{2}-\left(\frac{L_{z}^{2}}{\mathrm{sin}^{2}{\theta}}-a^{2}E^{2}\right)\mathrm{cos}^{2}\theta\\&-\left(L_{z}-aE\right)^{2}-\Delta\left(\frac{\partial {\mathcal S}}{\partial r}\right)^{2}-\left(\frac{\partial {\mathcal S}}{\partial \theta}\right)^{2}
\end{aligned}
\end{equation}
which can be rewritten as
\begin{equation}\label{eq-qq}
\begin{aligned}
&\Delta\left(\frac{\partial {\mathcal S}}{\partial r}\right)^{2}-\frac{1}{\Delta}\left[\left(r^{2}+a^{2}\right)E-aL_{z}\right]^{2}+\left(L_{z}-aE\right)^{2}+\delta r^{2}\\&=-\left(\frac{\partial {\mathcal S}}{\partial \theta}\right)^{2}-\left(\frac{L_{z}^{2}}{\mathrm{sin}^{2}{\theta}}-a^{2}E^{2}+\delta a^{2}\right)\mathrm{cos}^{2}\theta
\end{aligned}
\end{equation}
In Eq.~(\ref{eq-qq}), the left hand side depends on $r$ only, while the right hand side depends on $\theta$ only. So they must be separately equal to a constant $\mathcal{Q}$. This is the so-called Carter constant. For the regular black hole solution, Eq.~(\ref{eq-qq}) becomes
\begin{equation}
\begin{aligned}\label{eq-q-r}
\delta\frac{\left(\Sigma+L^{2}\right)^{4}}{\Sigma^{3}}&=\frac{1}{\Delta}\left[\left(r^{2}+a^{2}\right)E-aL_{z}\right]^{2}\\&
-\left(\frac{L_{z}^{2}}{\mathrm{sin}^{2}{\theta}}-a^{2}E^{2}\right)\mathrm{cos}^{2}\theta\\&-\left(L_{z}-aE\right)^{2}-\Delta\left(\frac{\partial {\mathcal S}}{\partial r}\right)^{2}-\left(\frac{\partial {\mathcal S}}{\partial \theta}\right)^{2} \, .
\end{aligned}
\end{equation}
For the singular one, Eq.~(\ref{eq-qq}) becomes
\begin{equation}\label{eq-q-s}
\begin{aligned}
\delta\frac{\left(\Sigma-L^{2}\right)^{3}}{\Sigma^{2}}&=\frac{1}{\Delta}\left[\left(r^{2}+a^{2}\right)E-aL_{z}\right]^{2}\\&
-\left(\frac{L_{z}^{2}}{\mathrm{sin}^{2}{\theta}}-a^{2}E^{2}\right)
\mathrm{cos}^{2}\theta\\&-\left(L_{z}-aE\right)^{2}-\Delta\left(\frac{\partial {\mathcal S}}{\partial r}\right)^{2}-\left(\frac{\partial {\mathcal S}}{\partial \theta}\right)^{2} \, .
\end{aligned}
\end{equation}
From Eq.~(\ref{eq-q-r}) and Eq.~(\ref{eq-q-s}), we see that in both spacetimes there is a Carter-like constant only when $L=0$ for massive particles, while for photons we have always a Carter-like constant.

\section{\label{s-4}radiative efficiency}

Geometrically thin accretion disks around black holes are normally described by the Novikov-Thorne model~\cite{nt-disk}. The accretion process can be approximated as follows (see, e.g., \cite{bambi2017black} and references therein for more details). The particles of the accreting gas slowly fall onto the central black hole by losing energy and angular momentum. When they reach the ISCO radius, they quickly plunge onto the black hole without emitting additional radiation. In general, the total power of the accretion process is $L_{acc}=\eta\dot Mc^{2}$, where $\eta=\eta_{r}+\eta_{k}$ is the total efficiency, $\eta_{r}$ is the radiative efficiency, and $\eta_{k}$ is the fraction of gravitational energy converted to kinetic energy of jets or outflows. The Novikov-Thorne model assumes that $\eta_{k}$ is negligible, and therefore the radiative efficiency of a Novikov-Thorne accretion disk is 
\begin{equation}
\eta_{NT}=1-E_{ISCO}
\end{equation}
where $E_{ISCO}$ is the energy per unit rest-mass of the gas at the ISCO radius.

\begin{figure*}
\begin{minipage}{0.48\linewidth}
\centerline{\includegraphics[width=8.5cm]{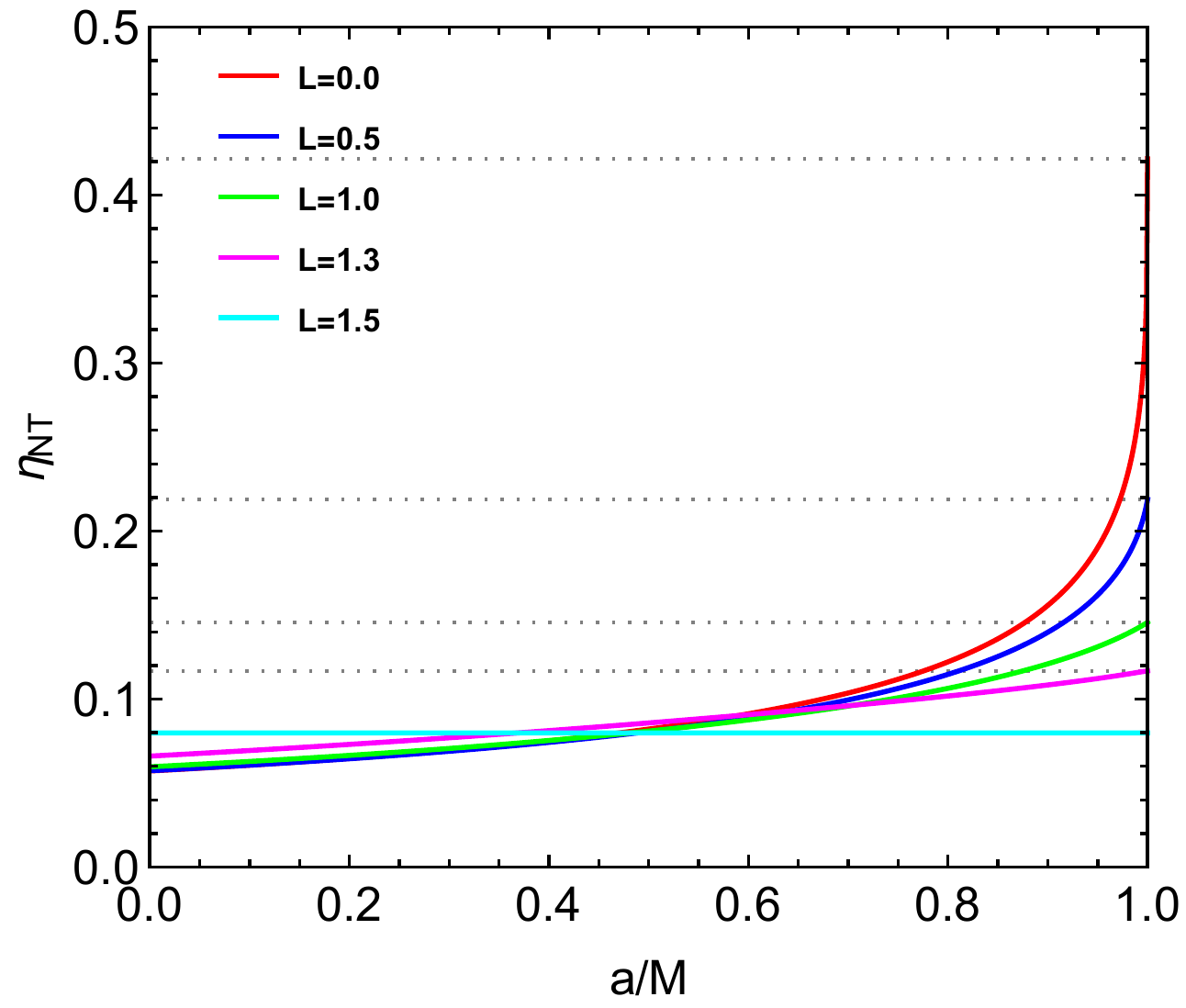}}
\centerline{(a)}
\end{minipage}
\hfill
\begin{minipage}{.48\linewidth}
\centerline{\includegraphics[width=8.5cm]{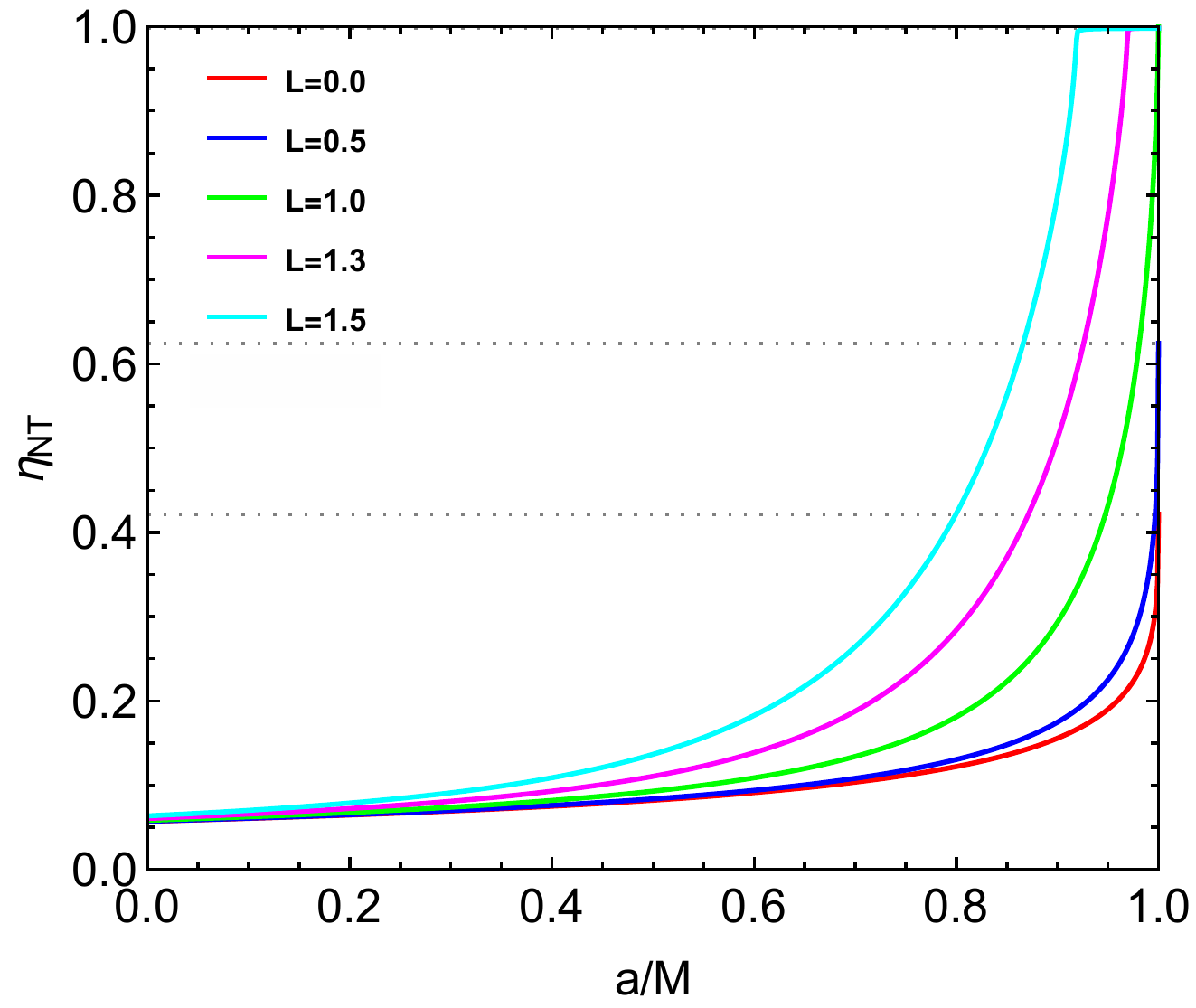}}
\centerline{(b)}
\end{minipage}
\vfill
\caption{Novikov-Thorne radiative efficiency $\eta_{NT}$ as a function of $a/M$ for a few different values of the parameter $L$ for (a) regular and (b) singular black hole spacetimes. \label{f-eta}}
\vspace{1.0cm}
\begin{minipage}{0.48\linewidth}
\centerline{\includegraphics[width=8.5cm]{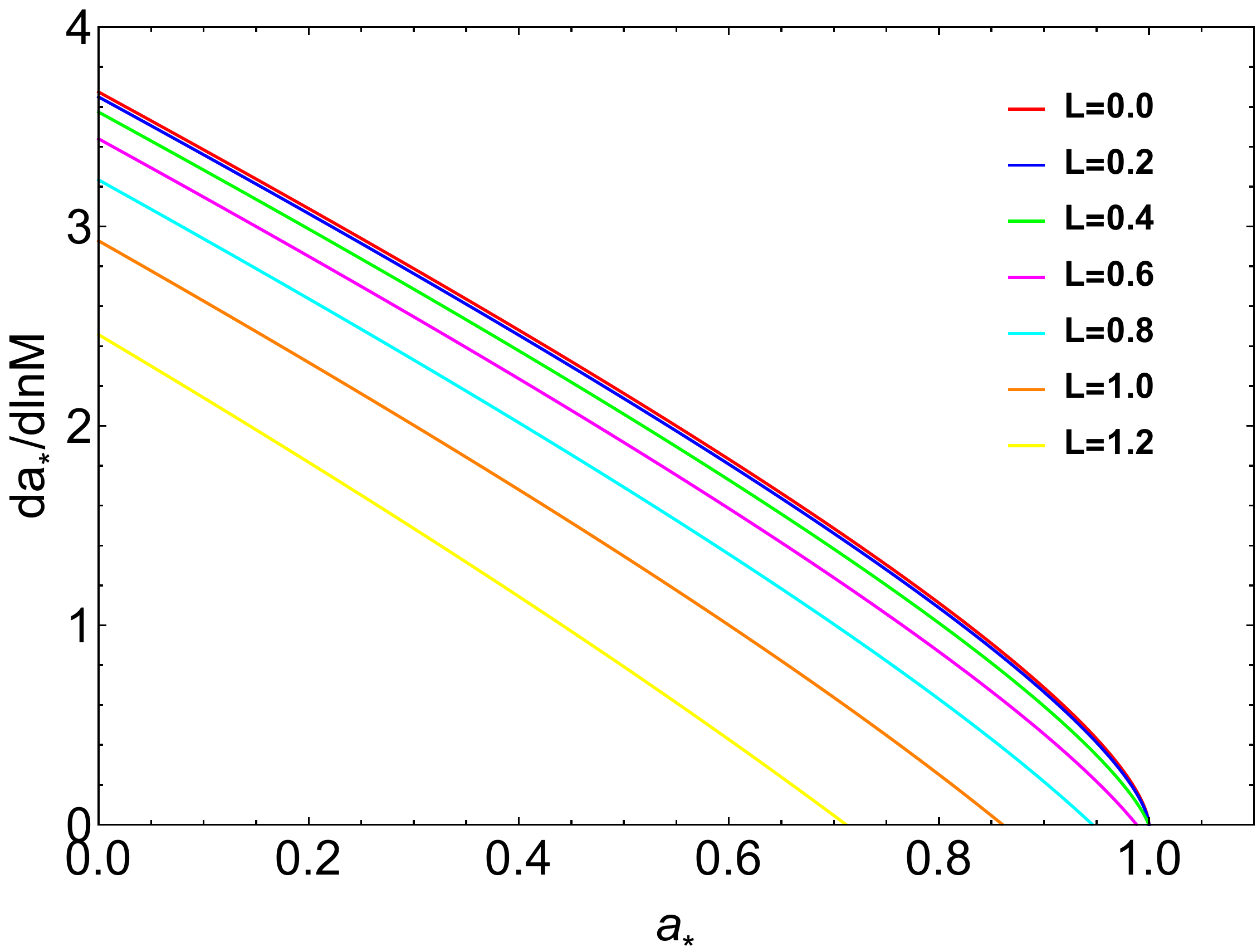}}
\centerline{(a)}
\end{minipage}
\hfill
\begin{minipage}{.48\linewidth}
\centerline{\includegraphics[width=8.5cm]{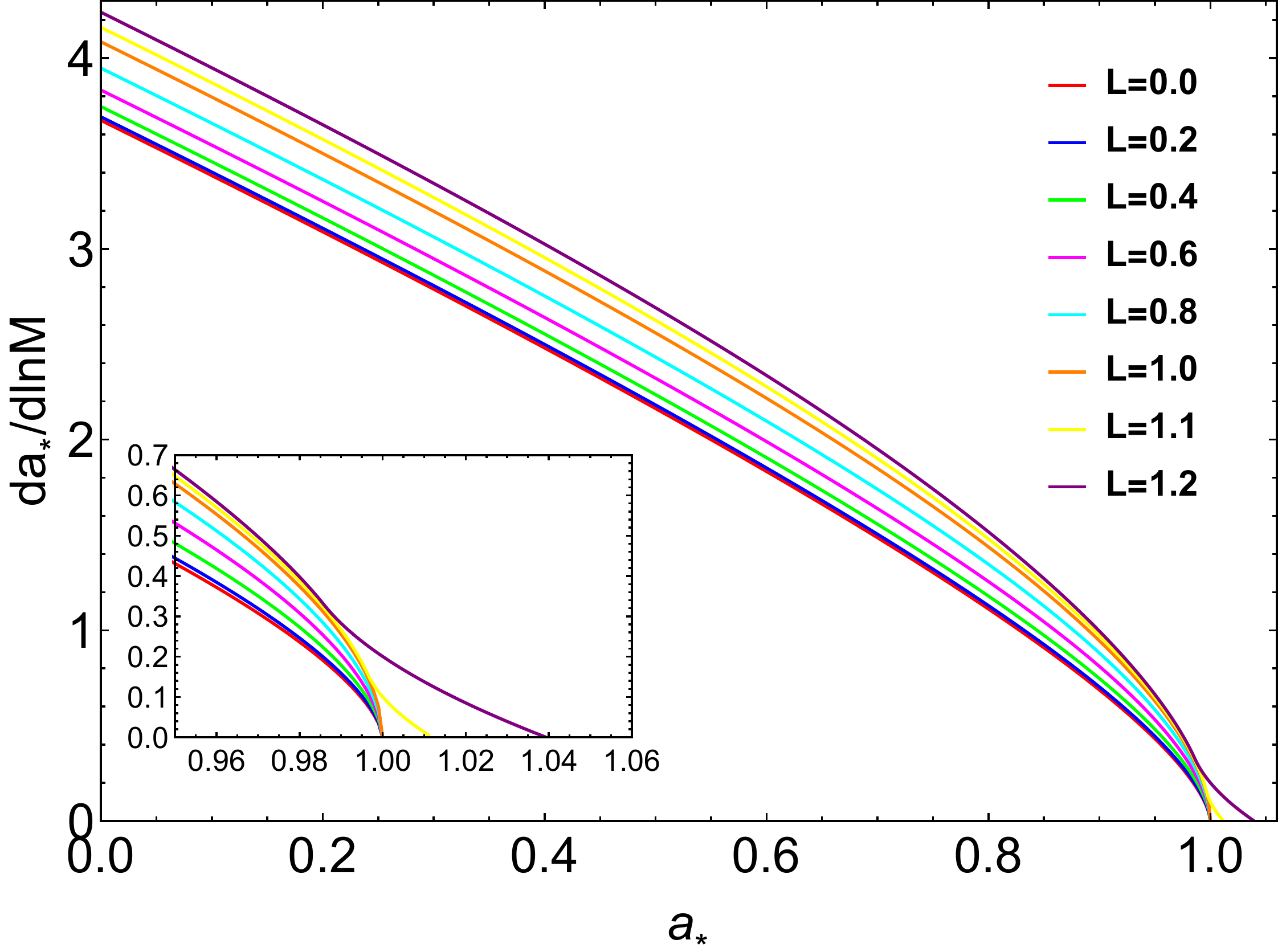}}
\centerline{(b)}
\end{minipage}
\vfill
\caption{$da_*/d\ln M$ as functions of the spin parameter $a_*$ for different values of the parameter $L$ for (a) regular and (b) singular black hole spacetimes. The inset in panel~(b) shows the details near $a_* = 1$. \label{f-spin}}
\end{figure*}

Fig.~\ref{f-eta} shows the Novikov-Thorne radiative efficiency $\eta_{NT}$ as a function of the spin parameter for different values of the parameter $L$ in regular and singular black hole spacetimes. It is interesting to note that the maximum radiative efficiency decreases as the value of the parameter $L$ increases for the regular spacetimes, and we have the opposite behavior for the singular ones. Since astronomical data suggest that $\eta_{NT} > 0.10$ is a conservative bound at least for some black holes~\cite{eta1,eta2,eta3}, we can obtain the observational constraint $L/M \lesssim 1.3$ for the regular black hole solutions. Such a very qualitative bound is consistent with the constraint $L/M < 1.2$ obtained from the iron line in~\cite{bambi2017testing}. In~\cite{Zhou:2018bxk}, we derived the constraint $L/M < 0.45$ from the analysis of the full reflection spectrum of the supermassive black hole in 1H0707--495, but the conformal factor studied was slightly different.

\section{\label{s-5}Destroying the event horizon}
The weak cosmic censorship conjecture asserts that the singularities produced in a gravitational collapse must be hidden behind an event horizon and cannot be seen by observers in the flat faraway region~\cite{penrose2002nuovo}. This is a conjecture and there is no proof, neither in Einstein's gravity nor in alternative theories of gravity. Several authors have studied the possibility of ``destroying''\footnote{The event horizon of a black hole cannot be destroyed by definition. Here we start from a stationary black hole spacetime and we ``introduce'' some small particles. The resulting spacetime may not be a black hole solution any longer. In practice, because of the complexity of the problem, the process is described by considering consecutive members of a family of stationary metrics and by checking if it is possible to ``jump'' from the spacetimes with black holes to those without black holes.} the event horizon of a black holes in order to create a naked singularity and find a violation of the weak cosmic censorship conjecture~\cite{rees1974black,hubeny1999overcharging,jacobson2009overspinning,gao2013destroying,toth2012test}. However, till now the conjecture seems to be correct. A different situation was discussed in Ref.~\cite{li2013destroying}, where it was shown that it is possible to destroy a singularity-free black hole. This is not a counterexample of the weak cosmic censorship conjecture because the destruction of the event horizon does not lead to any naked singularity.

Here, we want to study the possibility of destroying the event horizon in our regular and singular black hole spacetimes. As pointed out before, the event horizon of these metrics is the same as in the Kerr solution and the critical value of the spin parameter separating black hole spacetimes and horizonless spacetimes is $a_*^{crit} = 1$.

We consider an accreting black hole from a thin accretion disk~\cite{thorne}. The equilibrium value of the spin parameter can be evaluated as follows~\cite{bambi2011evolution}. The disk is in the equatorial plane and the disk's gas moves on nearly geodesic circular orbits. The gas loses energy and angular momentum and thus falls onto the central objects. When it reaches the ISCO, it plunges onto the massive body. If the gas is absorbed by the compact object,  with no further emission of radiation, the compact object changes its mass by $\delta M = \epsilon_{ISCO}\delta m$ and its spin angular momentum by $\delta J = \lambda_{ISCO}\delta m$, where $\epsilon_{ISCO} $ and $ \lambda_{ISCO}$ are, respectively, the specific energy and the specific angular momentum of the gas particle at the ISCO, while $\delta m$ is the gas rest-mass. The evolution of the spin parameter turns out to be governed by the following equation~\cite{bambi2011spinning}
\begin{equation}
\frac{da_{*}}{d \ln M} = \frac{1}{M}\frac{\lambda_{ISCO}}{\epsilon_{ISCO}} - 2a_{*}.
\end{equation}
When $da_{*}/d \ln M > 0$ , the accretion process spins the black hole up. When $da_{*}/d \ln M < 0$, the black hole is spun down. The equilibrium spin parameter is thus the one for which $da_{*}/d \ln M = 0$~\cite{bambi2011evolution}.

In Fig.~\ref{f-spin}, we show $da_{*}/d \ln M$ as a function of the spin parameter $a_{*}$ for $L$ from 0 to 1.2 in the case of the regular (a) and singular (b) spacetimes. The value of the equilibrium spin parameter is lower than 1 for regular black holes, and decreases as $L$ increases. We have also tried other regular rescaled metrics with different conformal factors, finding the same result.

For the singular black hole spacetimes, we find that the equilibrium spin parameter is greater than 1 when $L/M > 1$. Note that for these spacetimes the singularity at $r = L$ becomes naked as soon as $r_{\rm H} < L$, which happens before $a_* > 1$. In this sense, panel~(b) in Fig.~\ref{f-spin} has to be interpreted as the evidence that we can create a naked singularity when $L/M > 1$, not that we can create configurations with spin $a_* > 1$, because once there is a naked singularity we do not really have the accretion process under control. Even for the singular black hole spacetimes we have tried other singular rescaled metrics. It turns out that, at least with our set-up of an accretion disk spinning the central object up, we can destroy the black hole only when the exponent in $\left(1 - L^2/\Sigma\right)$ is odd, while we fail when the exponent is an even number.

\section{\label{s-6} Concluding remarks}

In the present paper, we have studied some general properties of two families of black hole spacetimes conformally equivalent to the Kerr solution. One of them was found in~\cite{bambi2017spacetime} and is regular everywhere, both in the sense it is geodesically complete and curvature scalars never diverge. The second spacetime introduced in this paper is singular, both in the sense it is geodesically incomplete and curvature scalars diverge. Together with the mass $M$ and the spin angular momentum $J$, these spacetimes are characterized by the presence of the parameter $L$, and for $L=0$ we exactly recover the Kerr solution.

We have studied the geodesic motion in these spacetimes, in particular equatorial circular orbits as their properties have more direct observational implications for astrophysical black holes accreting from thin disks. We have calculated the expected Novikov-Thorne radiative efficiency of a putative accretion disk around a similar black hole. Imposing the (conservative) observational constraint $\eta_{NT} > 0.1$, we find that $L/M < 1.3$ for the regular black hole spacetimes while there is no constraint for the singular ones.

Lastly, we have studied the possibility of ``destroying'' the event horizon in these spacetimes by overspinning the black hole with the matter in the accretion disk. Within our set-up we are unable to destroy the event horizon in the regular solutions, while we succeed in our attempt in the case of the singular ones (at least when the exponent in the conformal factor is odd). We may interpret our result either as the fact the cosmic censorship conjecture does not hold in the broken phase of conformally invariant theories of gravity or as the fact the cosmic censorship conjecture serves as a selection criterion to choose a ``good'' vacuum when the conformal symmetry gets broken.

\bibliographystyle{unsrt} 
\bibliography{ref}

\end{document}